\documentclass[conference]{IEEEtran}
\IEEEoverridecommandlockouts
\usepackage{cite}
\usepackage{amsmath,amssymb,amsfonts}
\usepackage{graphicx}
\usepackage{subfigure}
\usepackage{textcomp}
\usepackage{xcolor}
\usepackage[english]{babel}
\usepackage[utf8]{inputenc}
\usepackage{algorithm}
\usepackage[noend]{algpseudocode}
\usepackage{subfigure}

\def\BibTeX{{\rm B\kern-.05em{\sc i\kern-.025em b}\kern-.08em
T\kern-.1667em\lower.7ex\hbox{E}\kern-.125emX}}
\begin{document}

    \title{ Power Control Based on Multi-Agent Deep Q Network for D2D Communication \\
    {\footnotesize}
    }

    \author{\IEEEauthorblockN{Shi Gengtian,Takashi Koshimizu, Megumi Saito, Pan Zhenni,Liu Jiang, Shigeru Shimamoto}
    \IEEEauthorblockA{\textit{Graduate School of Fundamental Science and Engineering} \\
    \textit{Waseda University, Tokyo, Japan}\\
    shigengtian@akane.waseda.jp}
    }

    \maketitle
    \begin{abstract}
    In device-to-device (D2D) communication under a cell with resource sharing mode the spectrum resource utilization of the system will be improved. However, if the interference generated by the D2D user is not controlled, the performance of the entire system and the quality of service (QOS) of the cellular user may be degraded. Power control is important because it helps to reduce interference in the system. In this paper, we propose a reinforcement learning algorithm for adaptive power control that helps reduce interference to increase system throughput.
    Simulation results show the proposed algorithm has better performance than traditional algorithm in LTE (Long Term Evolution). 
\end{abstract}
    \section{Introduction}
With the emergence of more and more mobile multimedia services,
mobile communication networks are moving toward lower energy consumption and greater resource utilization and higher network capacity development. Among the device-to-device D2D communication \cite{b1}, the data between adjacent mobile terminals do not need to be forwarded by the base station but allows the local link communication to be established directly using the cellular network spectrum resources under the control of the base station. This flexible communication method alleviates the problem of base station processing bottleneck and coverage blind spot, the load bottleneck of the relay forwarding of the base station is alleviated, so that the mobile terminal has smaller transmission delay and less energy consumption\cite{b2}. 

At present, D2D works mainly in two ways: overlay mode and underlay mode \cite{b3}. The former means that the D2D link uses resources orthogonal to the original cellular user link spectrum resources. There is no interference between the D2D user and the cellular user.
However, due to the uncertainty of the number of D2D users, the spectrum resource utilization of the system is low. The signaling overhead is large\cite{b4}. The latter refers to the D2D link multiplexing the spectrum resources of the original cellular users. Current research focuses on underlay mode \cite{b5}.

\subsection{Related works}
\begin{figure}
    \centering
    \subfigure[D2D reuse downlink resource]{
    \begin{minipage}[b]{0.3\textwidth}
        \includegraphics[width=1\textwidth]{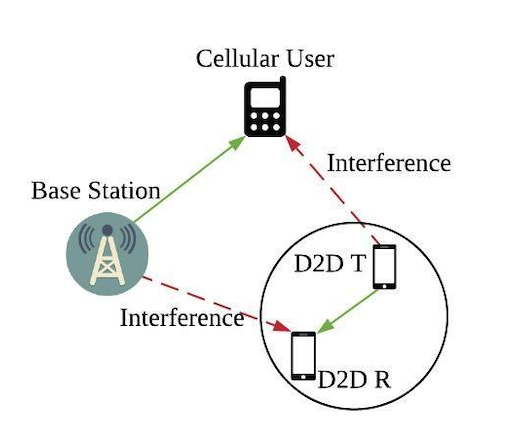}
    \end{minipage}
    }
    \subfigure[D2D reuse uplink resource]{
    \begin{minipage}[b]{0.3\textwidth}
        \includegraphics[width=1\textwidth]{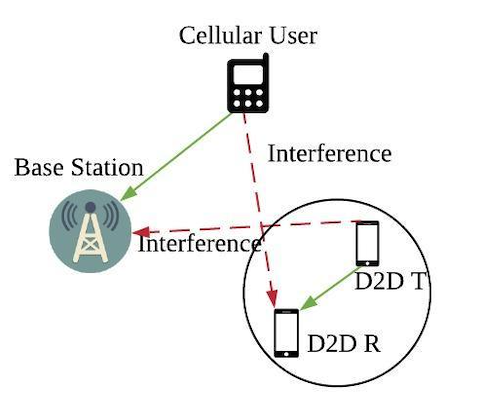}
    \end{minipage}
    }
    \caption{ D2D communication multiplexing cellular system resources.}
    \label{fig:intro}
\end{figure}

To reduce the co-channel interference caused by D2D multiplexing cellular users' resources in underlay mode, there are many related kinds of research, among which power control remains a hot research topic. D2D communication can multiplex cellular communication up link or down link resources. In both cases, D2D users interfere with cellular systems (D2C) and cellular systems interfere with D2D users (C2D) as shown in Fig.\ref{fig:intro}. D2D T is D2D transmitter, D2D R is D2D receiver.

When D2D communication reuse cellular communication up link resources,
the system can limit the power of the D2D transmitter. At the same time,
because the interference processing capability of the base station is strong,
D2C interference is easy to control and the system communication quality can be
guaranteed. \cite{b6} has studied the D2D user power optimization scheme for maximizing rate,
and derived the probability density function integral form of D2D user signal to interference and
noise ratio after power control.The down link interference of the base station to the D2D receiver can be effectively controlled by using the base station transmit signal beamforming technology \cite{b7}.

When D2D communication multiplexes cellular communication up link resources,
the system can limit the power of the D2D transmitter to ensure the communication quality of the cellular users.
There is also some interference between D2D users.
\cite{b8} has studied multiple user cellular users share the up link resources with a pair of D2D user pairs,
and reduce the interference of D2D to the cellular network through the receiving end area policy.
\cite{b9} and \cite{b10} has studied power control algorithm  game theory and stochastic geometry. Recent interest has focused on using machine learning methods to solve problems in wireless communications. Reinforcement learning is a powerful algorithm of machine learning. Q Learning is the basic algorithm of Reinforcement learning. Deep Q learning is an algorithm based on Q learning. In \cite{b11} has studied dynamic power allocation among femtocells
to deal with interference in the cellular network.
\subsection{Contribution}
This paper proposes a Deep Q Network \cite{b12} power control algorithm based on reinforcement learning to maximize system throughput while preserving the quality of service (QOS) of cellular users. The main contributions of this paper are two-folds:
\begin{itemize}
    \item Simplify the maintenance of Q table by using neural network approximate Q-Value.
    \item We compare the D2D power control algorithm based on Deep Q Network with Open Loop Power Control and Max Power Control and improve system throughput under the premise of ensuring communication quality of cellular users.
\end{itemize}

The rest of this paper is organized as follows. 
Section II describes the model of the system and formulates the problem. 
Section III introduces Deep Q Network for power control, 
Section IV presents simulation results and analysis. 
Section V concludes this paper.
    \section{System model and problem formulation
}

\subsection{System Model}
We consider the scenario where D2D users coexist with a set cellular user equipment(CUE) denoted by C,
C = \{1, 2, 3, …, C\} and a set of D2D users denoted by D, 
D = \{1, 2, 3, …, D\},
CUE and D2D users
are distributed uniformly at random with the coverage area of the base station. As Fig 3 shows, blue point is CUE, orange point is D2D transmitter, green point is D2D receiver, blue line means D2D communication.

We assume that cellular network operates in (Orthogonal Frequency Division Multiplexing) OFDM, there is no interference between CUE. In order to improve the spectrum utilization, D2D communication adopts the underlay-based working mode, that is, the D2D user reuses the used physical resource block (RB) of the CUE. It is further assumed that each pair of D2D user can only reuse at most one cellular RB.
\begin{figure}[htbp]
    \centerline{\includegraphics[scale=0.8]{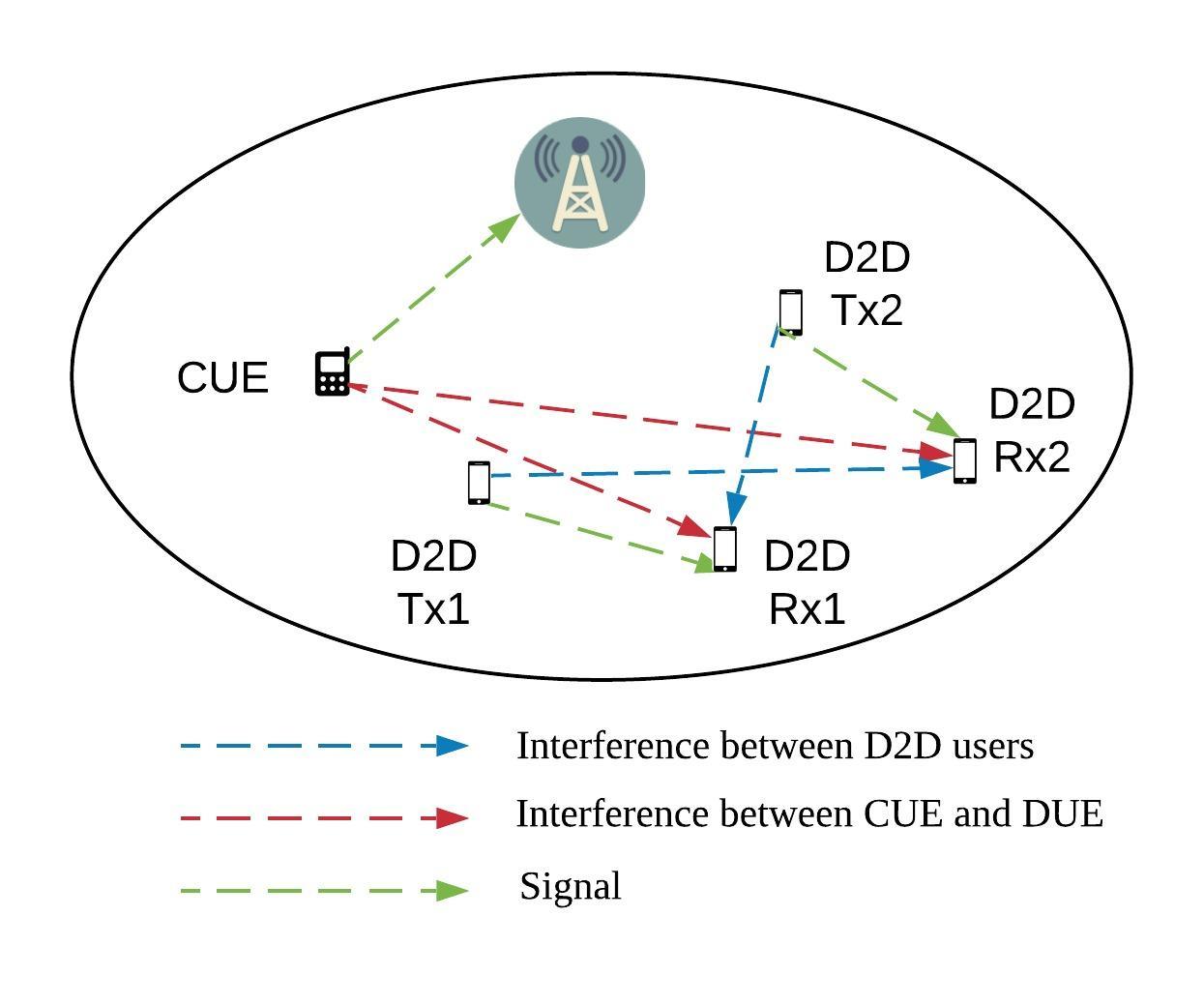}}
    \caption{ D2D communication in cellular networks.}
    \label{fig:system}
\end{figure}

Signal-to-interference-plus-noise-ratio (SINR) is a commonly used metric for the evaluation of communication quality in wireless networks. As illustrated in Fig\ref{fig:system}. We express the D2D user SINR on the $r$th RB as:
\begin{equation}
    \gamma_r^{D} = \frac{p_r^{d} \cdot G_r^d}{\sigma^2 + p_r^c \cdot G_r^{c} + \sum_{i=1}^{n} \cdot x_{i,k}  \cdot p_i \cdot G_{i,k}}
\end{equation}

Where $p_r^{d}$
and  $p_r^c$ denoted by D2D-Tx and CUE transmit power on the $r$th resource block respectively. $\sigma^2$ is the noise variance[2]. $x_{i,k}$ is DUE resource reuse factor, take the value 0, or 1 which is:

\begin{equation}
    x_{i,k} = \left\{\begin{matrix}
                         0 & \mbox{D2D pairs reuse CUE } r \mbox{th RB}    \\
                         1 &  \mbox{D2D pairs don't reuse CUE } r \mbox{th RB}
    \end{matrix}\right.
\end{equation}
Likewise, the SINR of CUE $c \in C$ is
\begin{equation}
    \gamma^{C} = \frac{p_r^c \cdot G_r^c}{\sigma^2 + \sum_{i=1}^{n} \cdot x_{i,k}  \cdot p_i \cdot G_{i,k} }
\end{equation}
$G$ is the gain from the transmitter to the receiver which can be expressed as follows:
\begin{equation}
    G = 10^{(-PL - shadowing)/10}
\end{equation}
PL is the path loss between the transmitter and receiver\cite{b4}.

\subsection{Formulation of problem}
The purpose of power control is to improve system performance including system capacity and D2D pair capacity. We assume and fix the transmit power of the cellular subscriber, then the power control for D2D can be achieved by follows
\begin{equation}
    max \sum_{r=1}^{R}{log_2(1+\gamma_r^C)} +log2(1+\gamma_r^D)
\end{equation}

\begin{figure}[htbp]
    \centerline{\includegraphics[scale=0.4]{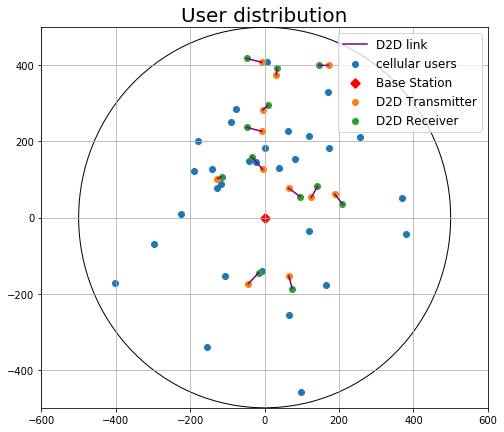}}
    \caption{ System Model}
    \label{fig:syste_model}
\end{figure}

It is obvious that $p_r^d$ is the objective function to maximize network capacity constraints. Increasing the transmit power of D2D will cause more interference to other D2D user pairs and cellular users, reduce the transmit power of D2D will reduce system throughput. To solve this problem, we proposed a method based on Reinforcement Learning to deal with it.

    \section{reinforcement learning based on power control algorithm}
In this section, we describe the basics of Reinforcement Learning, followed by demonstrate studying and propose a Deep Q Network Detailed solution. The reward feature allows agents to learn independently using only local information.

\subsection{Reinforcement learning}
\begin{figure}[htbp]
    \centerline{\includegraphics[scale=0.45]{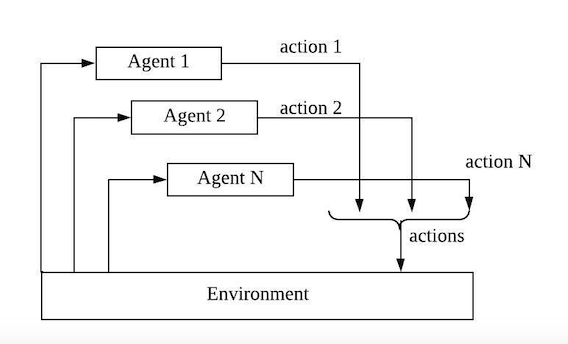}}
    \caption{Reinforcement Learning model}
    \label{fig:markov}
\end{figure}

As shown in Fig.\ref{fig:markov}, Reinforcement Learning consists of four parts: agent, state, action, and reward. There is no need for prior knowledge about the environment. The agent interacts with the environment by taking action and is rewarded for experience, and then the agent takes action through previously acquired experience traced by Q-Value. It is obtained according to the reward function, the agent takes action in the state then the state is transfer to another state, the reward of the action is obtained. Reinforcement learning is an algorithm based on Markov Decision Processing (MDP).

In our system we consider Agent, State, Action and Reward as follows:

Agent: Each D2D pair transmitter. 

State: The state is defined as the SINR of each cellular user on $r$th resource block at time $t$ which denoted by $S_t$ We assume that cellular user report the SINR to BS.
BS send the information to each D2D transmitter.

Action:The action consists of the transmit power level of each D2D defined as
\begin{equation}
    A = (a_r^1,a_r^2,a_r^3,...,a_r^p)
\end{equation}
where $\gamma$ and $p$ are denoted by on $\gamma$th RB and D2D user transmitting power level. In this paper, we use the $\epsilon$-greedy policy to generate the action.

Reward function: The reward function reflects the learning objectives of Reinforcement Learning, and our goal is to maximize system throughput. so we define reward function as
\begin{equation}
    \Re =\left\{\begin{matrix}
                    log_2(1+\gamma_r^c) + log_2(1+\gamma_r^d)& \textrm{if } \gamma_r^c >  \tau \\
                    -1
                    & \textrm{if } \gamma_r^c < \tau
    \end{matrix}\right.
\end{equation}

where $\gamma_r^c$ and  $\gamma_r^d$ are the SINR of cellular users and D2D users respectively, the on the $\gamma$th resource block.
In order to guarantee the SINR of the cell user's service quality performance, we set the SINR threshold ($\tau$) of the cellular user.

The Q-value can be obtained by the reward value obtained by the agent taking the action, and the relationship between the Q-value and the reward is
\begin{equation}
    Q_t(S_t, a_t) = R + \alpha Q_{t+1}(S_{t+1}, a_{t+1})
\end{equation}
where $\alpha$ is the attenuation factor ranging from 0 to 1. $R$ is reward. A high attenuation factor indicates that the correlation between the reward received by the agent and the future reward value is higher. The goal of our reinforcement learning is to let agents find the best strategy ($\pi$). By finding the action with the largest Q-value at state $t$. It can be expressed as
\begin{equation}
    \pi = max Q_t(S_t,a_t)
\end{equation}
The classic Q-learning needs to maintain the q table, update the Q-value and query the optimal strategy by Q-value. If the state-action space becomes huge, the maintenance of the Q-table will become difficult. To cope with this problem, Deep Q Network will be introduced.

\subsection{Deep Q Network}
\begin{figure}[htbp]
    \centerline{\includegraphics[scale=0.5]{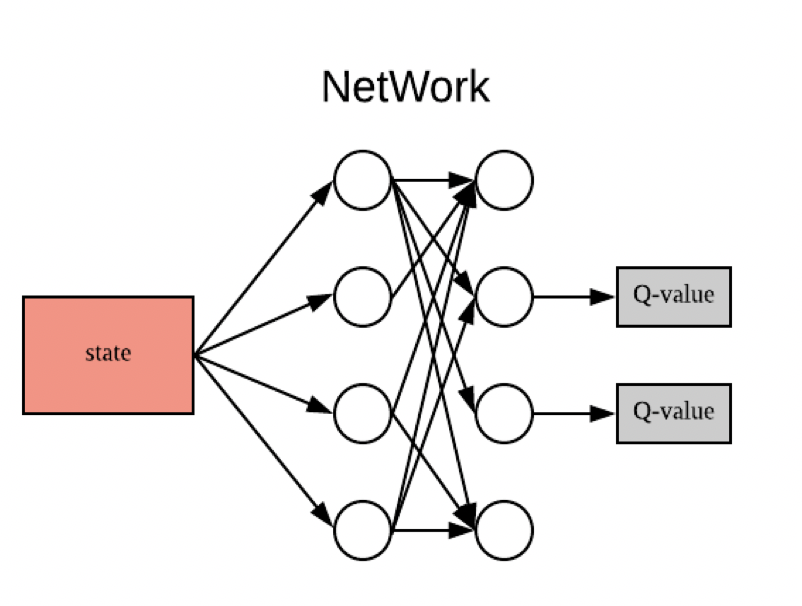}}
    \caption{Deep Q Network}
    \label{fig}
\end{figure}

Deep Q Network is a method of combining neural networks with Q-learning. As shown in Fig.\ref{fig}, State is the input of the neural network, the q-value corresponding to each action is the output of the neural network. The neural network is used as an approximation to find the best behavior value function. It can be expressed as
\begin{equation}
    Q(S_t,a_t;\Theta) \approx Q(S_t,a_t)
\end{equation}
$\theta$ is the neural network weight. $a_t$ is power level is selected by each D2D transmitter on each resource block.

In the iterative process, the neural network minimizes the following loss function by updating the weights by Adam \cite{b13}.
\begin{equation}
    Loss(\theta) = \frac{1}{n}\sum_{a_t=1}^{n}(y- Q(S_t,a_t))^2
\end{equation}
where
\begin{equation}
    y = r_t + maxQ(S_t, a_t)
\end{equation}
$r_t$ is the corresponding reward. 

Deep Q Network has a memory library for storing previous experiences. In each learning process, random extraction of previous energy can disrupt the correlation between experiences. In the problem of D2D power control, each iteration of state, action, reward will be stored in the memory.

\subsection{Training and Testing Algorithms}
The method we propose is the same as most machine learning algorithms, which is divided into training and testing. The data is generated by the agent interacting with the environment. Each training is used to optimize the weight of the DQN. The environment includes the transmit power of the CUE and the location information of the CUE and D2D are randomly generated. Deep Q learning with experience replay during the training phase. Training data is generated and stored in memory. For example, in the algorithm training, the agent is rewarded by selecting the level of random selection power in each iteration, and gradually increases and updates the weight of the neural network as the number of training increases. As shown in Algorithm 2, in the test phase, D2D is evaluated by selecting the power level corresponding to the Q value calculated by the neural network.

\begin{algorithm}
    \caption{Training stage}
    \textbf{INPUT: each $s \in S$ } \\
    \textbf{OUTPUT: trained Deep Q Network weight}\\
    \textbf{Initialize: Neural network architecture}\\
    \textbf{Initialize: Random weights $\theta$}
    \begin{algorithmic}[1] 
        \For{episode = 1, M}
        \For{$d$ = 1, D}
        \State Generate random value $x \in (0,1)$
        \If{$x < \varepsilon$}
        \State Chose random action
        \Else
        \State Chose $max Q_t(S_t,a_t)$ action
        \EndIf
        \State Calculate reward
        \State Calculate loss function $Loss$
        \State Back propagation 
        \State Update $\theta$ to minimize $Loss$
        \EndFor
        \EndFor\\
    \textbf{Return: Trained weight $\theta$}
    \end{algorithmic}
\end{algorithm}

\begin{algorithm}
    \caption{Testing stage}
    \textbf{INPUT: each $s \in S$, trained DQN weight $\theta$} \\
    \textbf{OUTPUT: Evaluation results}
    \begin{algorithmic}[1] 
        \For{$d$ = 1, D}
        \State Chose $max Q(S_t,a_t)$ action
        \EndFor
        \State Calculate D2D throughput
        \State Calculate system throughput\\
    \textbf{Return: Evaluation results}
    \end{algorithmic}
\end{algorithm}
    \section{Simulation and analysis}
\begin{figure}[htbp]
    \centerline{\includegraphics[scale=0.45]{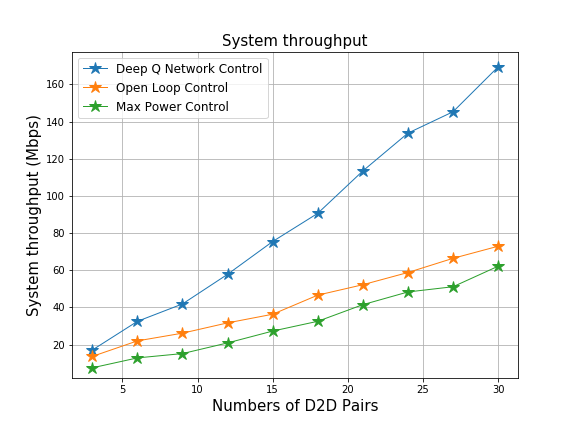}}
    \caption{System throughput as a function of the number of D2D users}
    \label{fig:System_throughtput}
\end{figure}

\begin{figure}[htbp]
    \centerline{\includegraphics[scale=0.45]{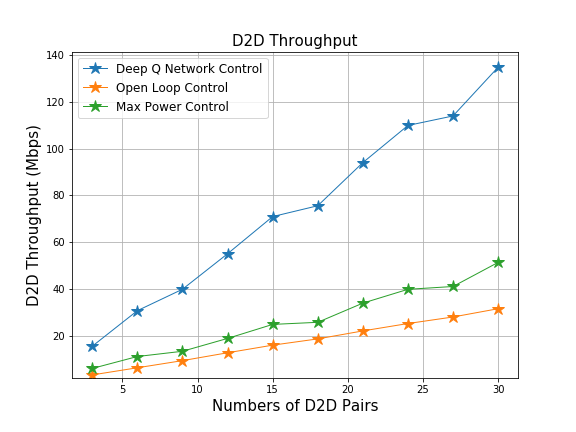}}
    \caption{D2D throughput as a function of the number of D2D users}
    \label{fig:D2D_throughput}
\end{figure}

In this section, we present simulation results to prove the performance for Deep Q Network with TensorFlow \cite{b14} .
We first describe simulation parameters and setting. In Fig\ref{fig:System_throughtput} and Fig\ref{fig:D2D_throughput}, we compare Deep Q network with Open Loop Power Control, Max Power Control method for LTE network in terms system throughput, the sum of D2D pairs throughput.

\subsection{NUMERICAL ANALYSIS}
We consider a system consisting of D2D pairs and 30 cellular users is uniformly distributed within the coverage of BS. The number of resource block equal to the number of cellular users is 30 The threshold of cellular users SINR is $\tau = 6db$. Every D2D user pair has been assigned one resource block, The learning rate of neural network is 0.001, The decay factor $\alpha$ is 0.95, probability $\varepsilon$ is 0.1. The layers of neural has 200 neurons, between the layers we use activate function Relu, The other parameters has shown in Table1.

\begin{table}[htbp]
    \caption{PARAMETERS OF SIMULATION}
    \begin{center}
        \begin{tabular}{|c||c|}
            \hline
            \textbf{\textit{Parameter}}& \textbf{\textit{Value}}\\
            \hline
            $p_{max}$& 23dbm\\
            \hline
            Thermal noise& -176dbm\\
            \hline
            Resource block bandwidth& 180kHz\\
            \hline
            pathloss model between BS and users& $15.3+36.6{log10(d(km))}$dB\\
            \hline
            pathloss model between users& $28+40{log10(d(km))}$dB\\
            \hline
            Macro BS antenna gain& 17dBi\\
            \hline
            User antenna gain& 4dBi\\
            \hline
        \end{tabular}
        \label{tab1}
    \end{center}
\end{table}

\subsection{Result and Analysis}

Fig $\ref{fig:System_throughtput}$ depicts the system throughput of the different algorithms. We can observe that as the number of D2D users increases, system throughput increases. Since we set the interference threshold of cellular users to ensure the communication quality of cellular users, and Deep Q Network has better adaptive ability to handle interference between D2D users. On We can observe obviously Deep Q Network has higher performance. In addition, as D2D users increase, The gap between Deep Q Network with Open Loop Power Control and Max Power Control becomes larger. 

Finally, Fig.$\ref{fig:D2D_throughput}$ depicts the total D2D throughput of different algorithms. It can be observed that Deep Q Network can reduce the interference between D2D to increase the total throughput of D2D users in the system and the performance is better than Open Loop Power Control and Max Power Control. With the increase of D2D users in the system, the performance advantage of Deep Q Network is more obvious.

\section{CONCLUSION}
This paper explores the optimization of power control in a single cell cellular network scenario where DUE and CUE coexist in underlay mode. Reinforcement learning is an appropriate method for adaptive power control, this study uses Deep Q Network for power control under the premise of ensuring the communication quality of cellular users, reducing interference and improving system throughput. Experimental results show that the algorithm has better performance than traditional algorithm.

\end{document}